\documentclass{article}

\usepackage{latexsym,amssymb,amsmath}
\usepackage{graphicx}

\def\GeV{{\rm\ GeV}}
\def\ve{\varepsilon}

\def\be{\begin{equation}}
\def\ee{\end{equation}}
\def\bea{\begin{eqnarray}}
\def\eea{\end{eqnarray}}

\def\ea{{\it et al.}}

\begin{document}
\title{Reanalysis of low-energy electron-proton scattering data and proton radius}
\author{Dmitry~Borisyuk$^1$, Alexander~Kobushkin$^{1,2}$\\[2mm]
\it\small $^1$Bogolyubov Institute for Theoretical Physics, 14-B Metrologicheskaya street, Kiev 03680, Ukraine\\
\it\small $^2$National Technical University of Ukraine "Igor Sikorsky KPI", 37 Prospect Peremogy, Kiev 03056, Ukraine}
\maketitle
\begin{abstract}We reanalyze electron-proton scattering data in the momentum transfer region $Q^2 < 1\GeV^2$, which were used to determine proton electric radius, with main focus on the Mainz experiment (Bernauer et al., 2010). We examine dependence of fit results and $\chi^2$ on the cut-off in $Q^2$ and degree of fitting polynomial, using pseudo-data and compare it with the case of real experimental data. We find that under some assumptions, the data could be consistent with the radius, obtained from muonic hydrogen.
\end{abstract}

\section{Introduction}

The proton radius puzzle originates in 2010, when the precise measurements of the Lamb shift in muonic hydrogen were performed,
allowing to extract the proton electric radius with unprecedented accuracy: $r_E = 0.84184(67)$~fm \cite{rE-muon}.
This value, which was further refined in 2013 as 0.84087(39)~fm \cite{rE-muon-2},
severely contradicted one obtained from a combination of past electron-proton scattering experiments and usual hydrogen levels, 0.8768(69)~fm \cite{rE-old-ep}.
Analyses of scattering data alone resulted in even higher values, such as 0.895~fm \cite{Sick}.

Further, large $ep$ scattering experiment at Mainz Microtron was held,
having taken about 1400 data points at $Q^2$ from 0.01 to 1 $\GeV^2$,
and yielded a result of 0.875(5)(4)(2)(5)~fm, which agreed with previous scattering data
and contradicted muon hydrogen results \cite{Bernauer}.

Various explanations of the discrepancy were proposed, including unaccounted QCD corrections to muonic hydrogen levels, effect of an undiscovered particle, or violation of the $e$-$\mu$ universality \cite{Rev1,Rev2}.

Recently, results of the PRad experiment at JLab were published,
where, using $ep$ scattering at very low momentum transfers,
the value of 0.831(7)(12)~fm was obtained \cite{PRad}, which is even lower than muonic results and agrees with them.
The question, however, still remains, why previous scattering experiments,
including Mainz experiment, gave so different results? We will try to find an answer.

Higinbotham \cite{Higinbotham} have found that scattering experiments are consistent with lower radius,
but analyzed only two experiments (except Mainz), and, importantly, restricted kinematics to very low-$Q^2$ data.
However, it is natural to expect that the more data points (i.e. more information) we are using
--- the more precise results will be obtained.
It is {\it a priori} unclear why discarding accurately measured data should lead to any improvement.
Specifically, as it was noted \cite{Sick}, cutting data off at too low $Q^2$ increases the effect of systematic errors (even if ``renormalization'' of data is performed), leading to unreliable results for radius.

In the present paper we will apply the fitting method, based on Ref.~\cite{ourRadius}, to the scattering data, and study the dependence of the results on different fit parameters.

\section{Fitting procedure}\label{Sec:Fitting}

We use basically the same fitting procedure as used in Ref.~\cite{ourRadius}. We minimize the following function:
\be \label{chi2}
 \chi^2 = \sum_{e,i}
  \left(
   \frac{\sigma^{\rm exp}_{e,i}-N_e \sigma^{\rm th}_{e,i}}{d\sigma_{e,i}}
  \right)^2
 + \sum_e \left( \frac{N_e-1}{dN_e} \right)^2,
\ee
where $e$ enumerates experiments, $i$ --- data points within each experiment,
$\sigma^{\rm exp}_{e,i}$ and $d\sigma_{e,i}$ are measured cross-sections
and their point-to-point errors, $N_e$ are normalization factors
(to be determined by fitting), $dN_e$ are normalization uncertainties
and $\sigma_{e,i}^{\rm th}$ is theoretical cross-section:
\be \label{sigma}
 \sigma_{e,i}^{\rm th} = \ve G_E^2(Q^2) + Q^2 G_M^2(Q^2)/4M^2 + \sigma^{2\gamma}(Q^2,\ve),
\ee
where $\sigma^{2\gamma}(Q^2,\ve)$ is two photon exchange correction,
calculated according to Refs.~\cite{ourDisp,ourPiN}.
The proton form factors (FFs) $G_E$ and $G_M$ are parameterized as follows:
\be
  G_E(Q^2) = (1-\xi/\xi_0)^2 \sum_{k=0}^{n_E} a_k \xi^k, \qquad
  G_M(Q^2) = (1-\xi/\xi_0)^2 \sum_{k=0}^{n_M} b_k \xi^k,
\ee
where $\xi = Q^2/(1+Q^2/\xi_0)$ with $\xi_0 = 0.71 \GeV^2$ (thus the multiplier in front of the sum is the dipole).
We two variants of the fit, where $n_M$ is equal either $n_E$ or $n_E-1$.
In both cases we call this ``fit of degree $n = (n_E+n_M)/2$'', that is, fit degree is $n_E$ for $n_M=n_E$ and $n_E-1/2$ for $n_M=n_E-1$.
The proper FF normalization is ensured by keeping $a_0 = G_E(0) = 1$ and $b_0 = G_M(0) = 2.793$ fixed.
After the fitting, $r_E$ is calculated via usual relation
\be\label{radii}
  r_E^2 = -\frac16 \left. \frac{dG_E}{dQ^2} \right|_{Q^2=0}
\ee

\section{Data}

We use two datasets:
\begin{itemize}
\item world data measured prior to 2010 (the same as used in Ref.~\cite{ourRadius},
hereafter referred to as ``old data''), which consists of 370 points for $Q^2 \le 1\GeV^2$, and
\item data of the Mainz experiment \cite{Bernauer,BernauerThesis} (``Bernauer data''), 1422 points.
\end{itemize}
In the latter case, the data consist of 34 series, each having its own normalization coefficient. In the original work, the normalization uncertainties were not estimated, but the corresponding coefficients were determined by fitting.
These 34 coefficients are ``entangled'' --- they are composed of 31 quantities, denoted $n_1 .. n_{31}$ in Ref.~\cite{BernauerThesis}.
To simplify the fitting, we exclude 3 smallest series (with coefficients $n_{14}$, $n_{18}$ and $n_2n_5$) for the coefficients of the rest to become independent.
Thus we are left with 1375 points in 31 series. When fitting, we treat these series as independent experiments, and the last term in Eq.~(\ref{chi2}) is dropped.


Before the fit, data are cut-off in $Q^2$, that is, only data with $Q^2 \le Q_{max}^2$ are selected, where $Q_{max}^2$ is a parameter, which we can change along with the fit degree $n$.

\section{$Q^2$ cut-off analysis}
\begin{figure}
\centering
\includegraphics[width=0.32\textwidth]{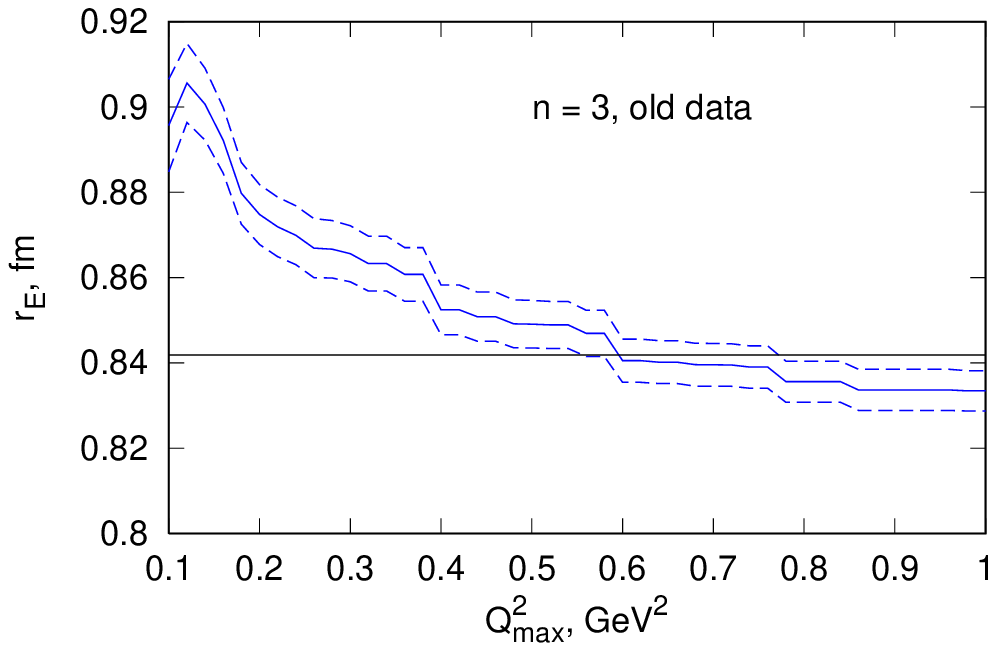}
\includegraphics[width=0.32\textwidth]{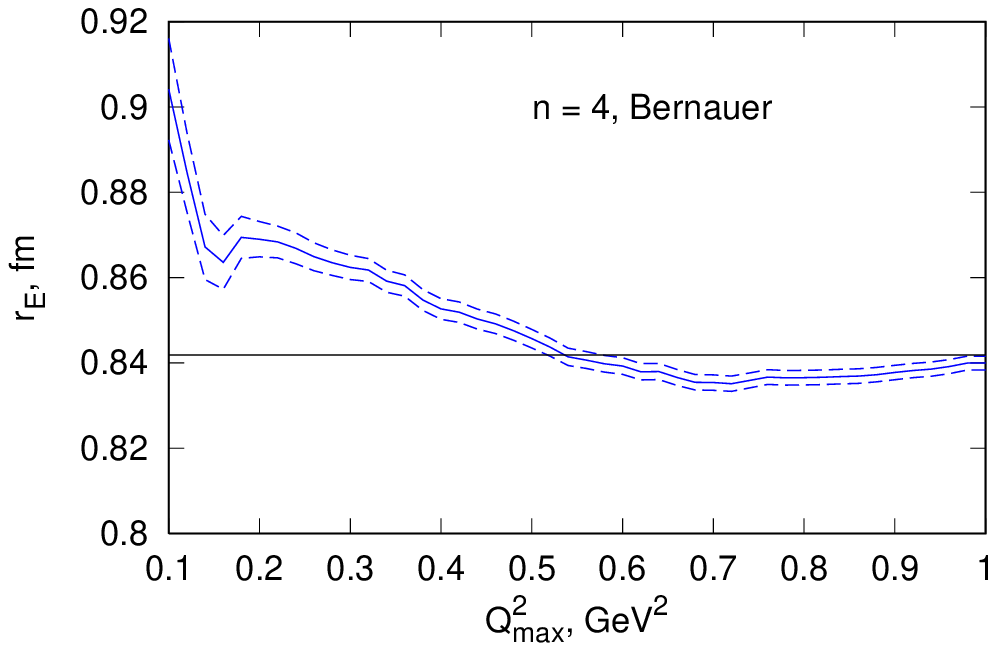}
\includegraphics[width=0.32\textwidth]{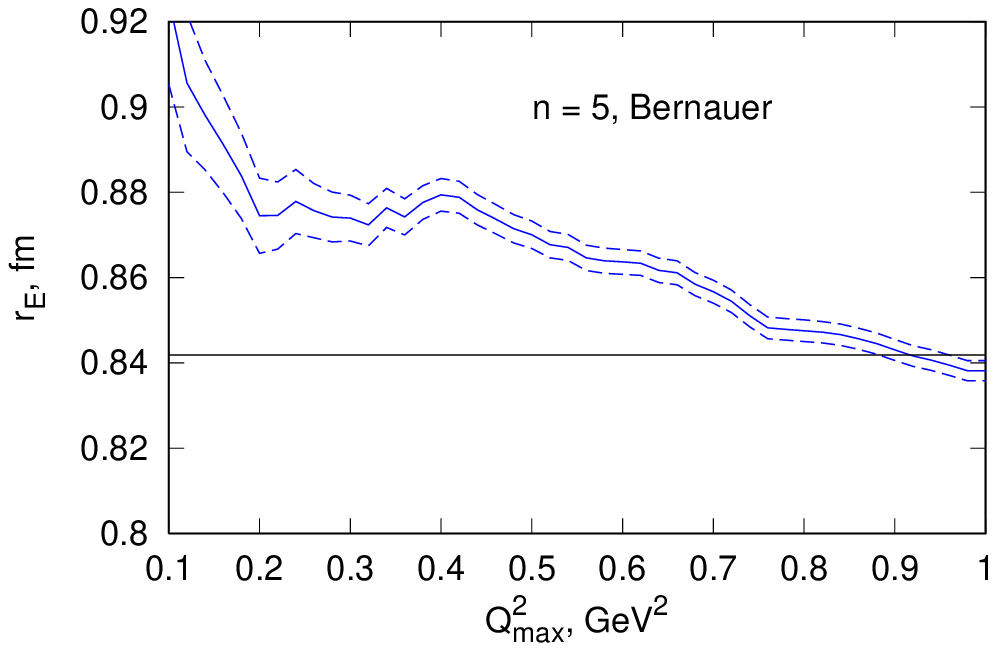}
\caption{Fit results: $r_E$ as a function of $Q_{max}^2$ with different $n$; horizontal line is muonic result.}
\label{fig:Qmax}
\end{figure}

In Ref.~\cite{ourRadius} we tried to determine optimal $Q^2_{max}$ by balancing between statistical error of $r_E$,
which decreases with $Q^2_{max}$, and systematic one, which increases.
But, as it was already noted there, the latter depends on the true FFs, which are unknown.
Thus, in Ref.~\cite{ourRadius} several realistic FF parameterizations were used to estimate systematic error instead of the true FFs. However such a result is still model dependent. Moreover, it is hard to guess which peculiarities of $Q^2$ dependence of the FFs contribute to the systematic error of the radius, and in which manner. When polynomials of high degree are involved, it is quite possible that some minor differences between true FF and model parameterization can cause significant change in the error. Thus in the present paper we avoid that approach.

At first, we just try to test the $Q^2_{max}$ dependence of the extracted radius.
Surprisingly, as we see in Fig.~\ref{fig:Qmax}, with increasing $Q^2_{max}$ the radius tends to a constant value,
which is very close to one obtained from muonic hydrogen.
This happens independently of $n$, and this seems not accidental, since the old data behave similarly.

As it was mentioned earlier, low cut-offs in $Q^2$ could lead to improper normalization of the data, since more and more points from each experiment are left out. As an exaggerated example, if the cutoff is such that only one data point from an experiment remains, then it becomes completely useless, since it could be ``renormalized'' to any value.

\begin{figure}
\centering
\includegraphics[width=0.32\textwidth]{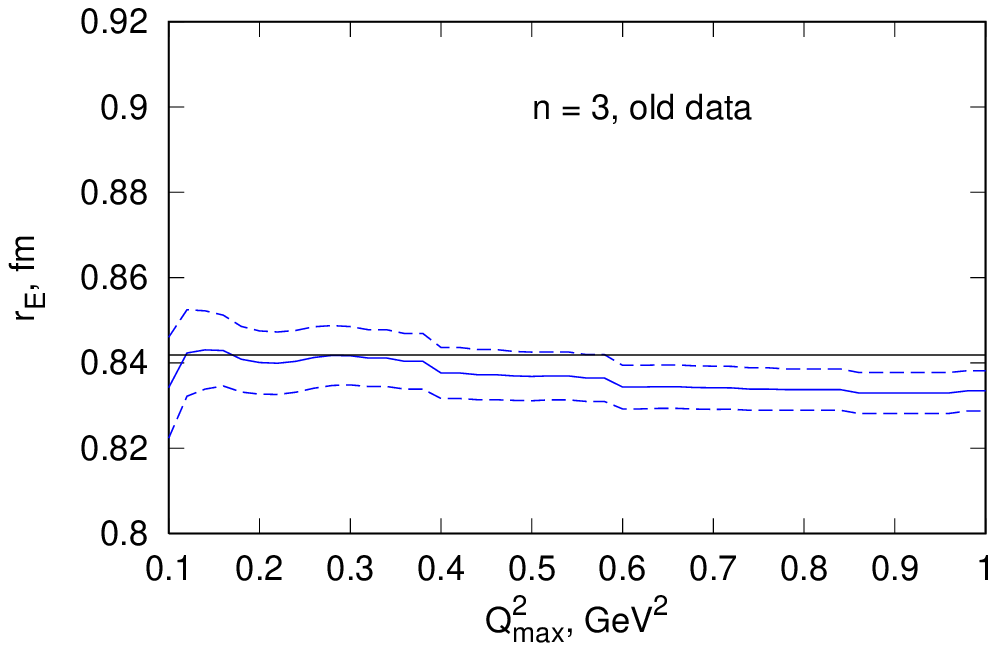}
\includegraphics[width=0.32\textwidth]{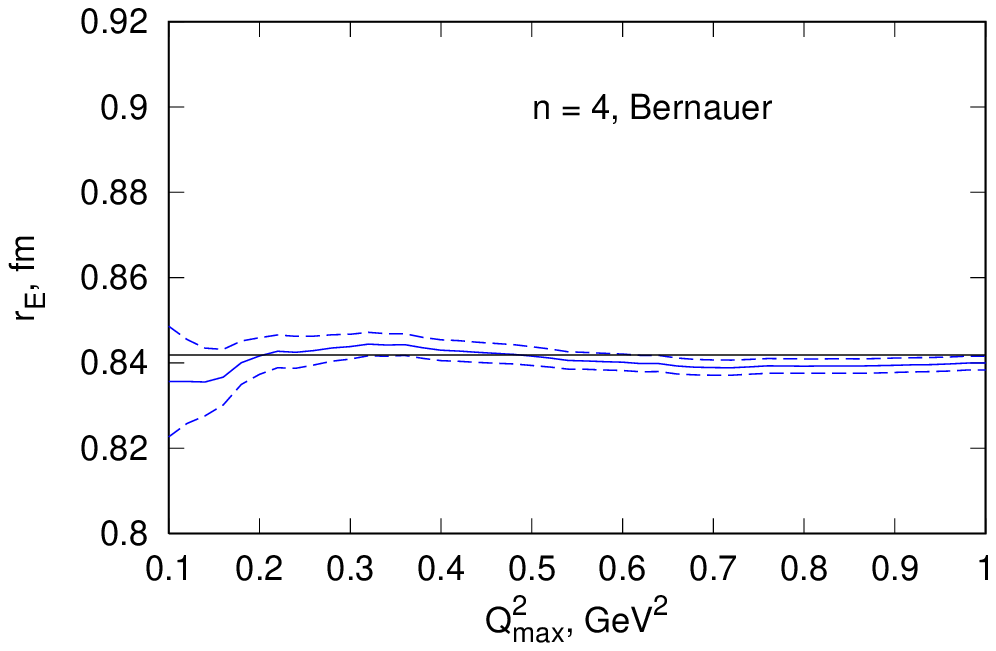}
\includegraphics[width=0.32\textwidth]{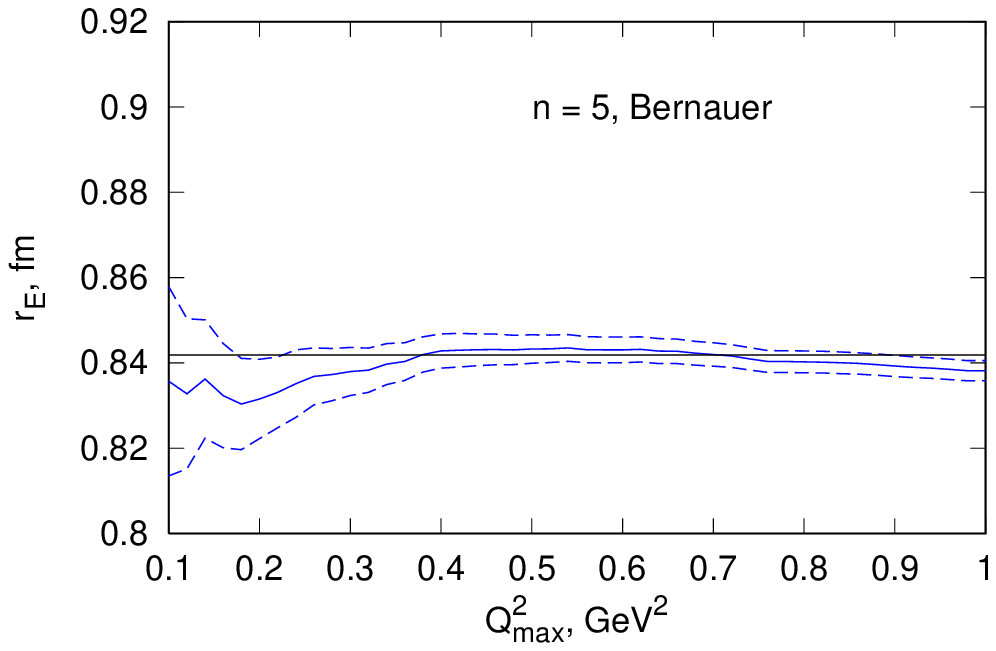}
\caption{Same as Fig.~\ref{fig:Qmax}, but normalization was always determined at $Q_{max}^2=1\GeV^2$.}
\label{fig:Qmax-n}
\end{figure}

To check the effect of the cut-off on normalization, we could do the following trick: first, the normalization coefficients are determined by fit with highest $Q^2_{max}$ ($1 \GeV^2$), and then they are fixed and the fit is repeated with varying $Q^2_{max}$.
The results are shown in Fig.~\ref{fig:Qmax-n}: is this case, the extracted $r_E$ becomes almost independent of $Q^2_{max}$ and still close to the muonic result.

The results obtained above suggest that to obtain proper normalization, it is better to use higher $Q_{max}^2$ (and corresponding optimal fit degree $n$). Now we are to choose optimal $n$.
As it was noted, the balance between statistical and systematic error is difficult to probe without knowing true FFs.
We will try to circumvent this problem by studying dependence of $\chi^2$ and extracted radius on $n$, using pseudodata.

\section{Pseudo-data simulation}

\begin{figure}
\centering
\includegraphics[width=0.32\textwidth]{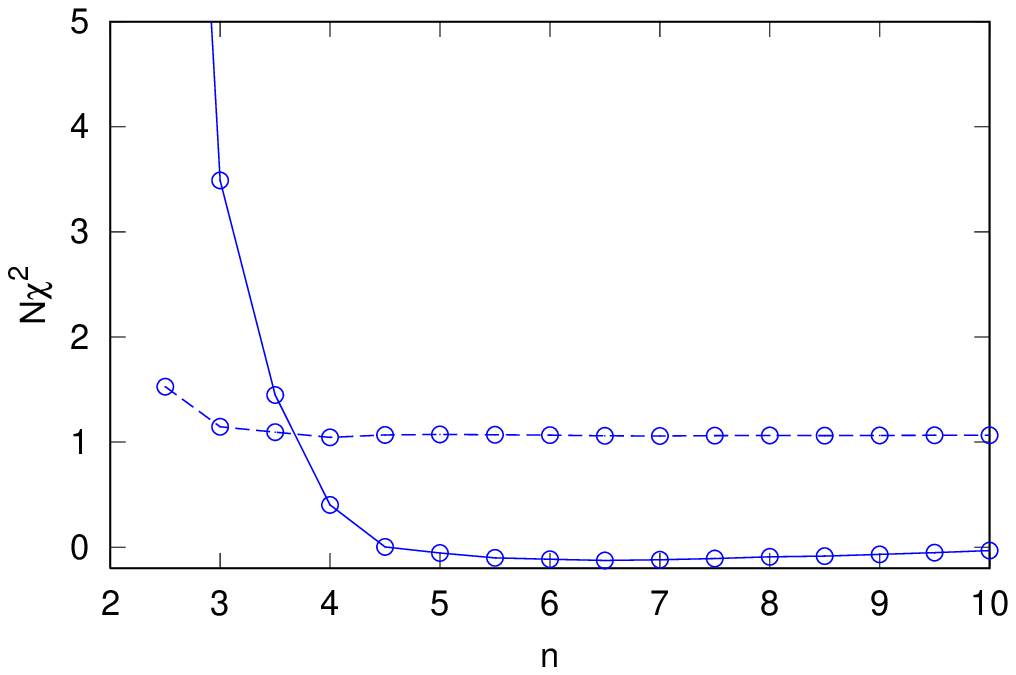}
\includegraphics[width=0.32\textwidth]{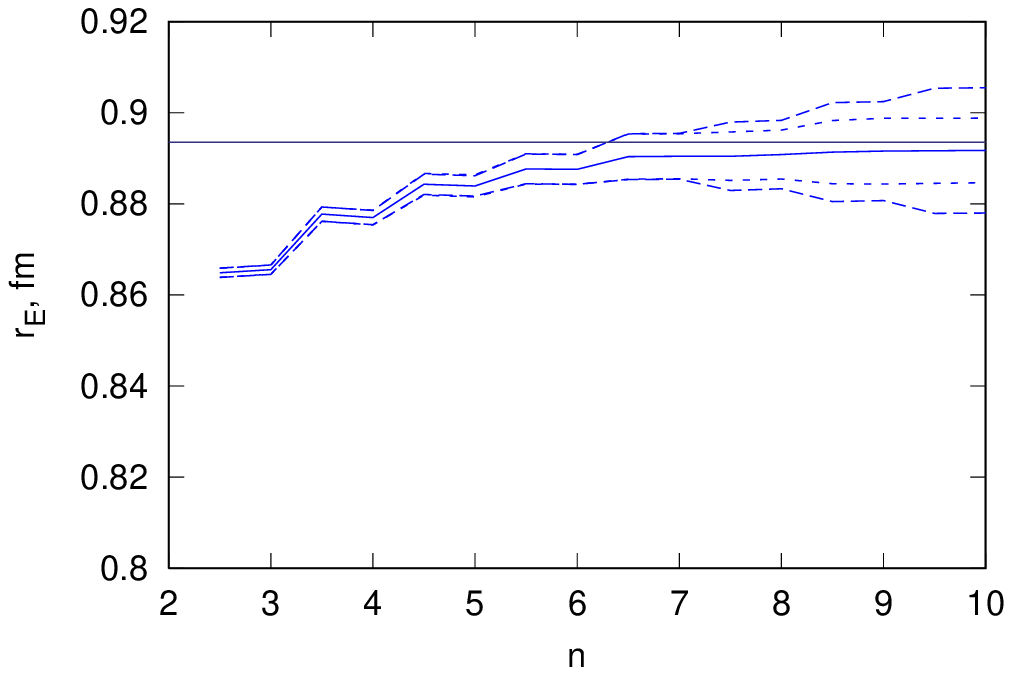}
\caption{Pseudo-data fitting results at $Q_{max}^2 = 1 \GeV^2$ and different $n$. Left: mean $N\chi^2$ (solid), and its standard deviation (dashed). Right: mean extracted radius (solid), with $\pm 1\sigma$ errors (dashed); horizontal line is true value.}
\label{fig:pseudo}
\end{figure}

The pseudodata was generated at the experimental data kinematics of Ref.~\cite{Bernauer} according to Eq.~(\ref{sigma})
with randomly generated point-to-point and normalization errors added:
\be
\sigma_{e,i}^{\rm pseudo} = 
 (1 + \zeta_e dN_e) \left[ \ve G_E^2(Q^2) + Q^2 G_M^2(Q^2)/4M^2 \right] + \zeta_{e,i} d\sigma_{e,i}
\ee
where $\zeta$ is random quantity with standard normal distribution (with zero mean and dispersion unity). Since the normalization uncertainties were not estimated in Ref.~\cite{Bernauer}, we arbitrarily set $dN_e = 2\%$.
The FF parameterization used was \cite{MMDFit}, as it has one of the most complicated forms among popular parameterizations.
The proton radius value, associated with this parameterization, is $r_E^{\rm th} = 0.893$ fm.
Then the procedure described in Sec.~\ref{Sec:Fitting} was applied to obtain $r_E$ and $\chi^2$ for different $n$ (of course, except the two photon exchange correction which is not needed here).
This pseudo-data extraction was repeated 400 times to collect sufficient statistics,
and the average $r_E$ and $\chi^2$ and their respective r.m.s. deviations were obtained.
The results are plotted in Fig.~\ref{fig:pseudo}.

Left panel of Fig.~\ref{fig:pseudo} shows ``normalized'' $\chi^2$ (further denoted $N\chi^2$) as a function of fit degree $n$.
Here ``normalized'' means that $\chi^2$ transformed so that, independently of number of degrees of freedom $d$, the resulting quantity is distributed according to standard normal distribution:
\be
N\chi^2 = {\rm normcdf}^{-1}({\rm chi2cdf}(\chi^2, d))
\ee
where chi2cdf and normcdf are cumulative distribution functions for $\chi^2$ and standard normal distributions, respectively.
For large $d$ this means
\be
N\chi^2 \approx \frac{\chi^2 - d}{\sqrt{2d}}
\ee
Such quantity is useful since it allows to intuitively inspect quality of the fit: say, $N\chi^2=3$ is as rare event as $3\sigma$ deviation, $N\chi^2=4$ is equivalent to $4\sigma$, etc.

The right panel shows extracted radius in comparison with true value $r_E^{\rm th}$ (obtained from parameterization). The dashed lines are $1\sigma$ error bounds: long-dashed --- mean estimated error in $r_E$, short-dashed --- standard deviation of $r_E$.
At small $n$ they coincide, but at large $n$ the latter becomes substantially smaller --- probably because the (linear) error estimate turns incorrect due to high nonlinearity of the fit.

Main conclusions from this pictures are:
\begin{itemize}
\item $N\chi^2$ decreases with $n$, and eventually stabilizes. This is because actual $\chi^2$ is the sum of two terms: the first, which is independent of point-to-point errors and sharply depends on $n$, reflecting how the true FF can fitted by the fit function, and the second, which depends on errors and is $\chi^2$-distributed. When the former becomes small, $N\chi^2$ stabilizes.
\item When $N\chi^2$ stabilizes, the extracted $r_E$ becomes close to true value within errors.
\item When stable, $N\chi^2$ is close to zero, and its standard deviation is close to one as expected from its definition.
\end{itemize}

We also repeated the above-described numerical experiment with two other FF parameterizations: \cite{ArringtonFit,KellyFit}. The results were essentially the same, except that in this case $N\chi^2$ stabilizes earlier: at $n \approx 4$ instead of $n \approx 6$ above. Most likely, this is because simpler functions, used in these parameterizations, need less terms of the polynomial to be described accurately.

Now let us look at the same plot for the real experimental data, keeping these properties in mind.

\section{Real data results and discussion}

\begin{figure}
\centering
\includegraphics[width=0.32\textwidth]{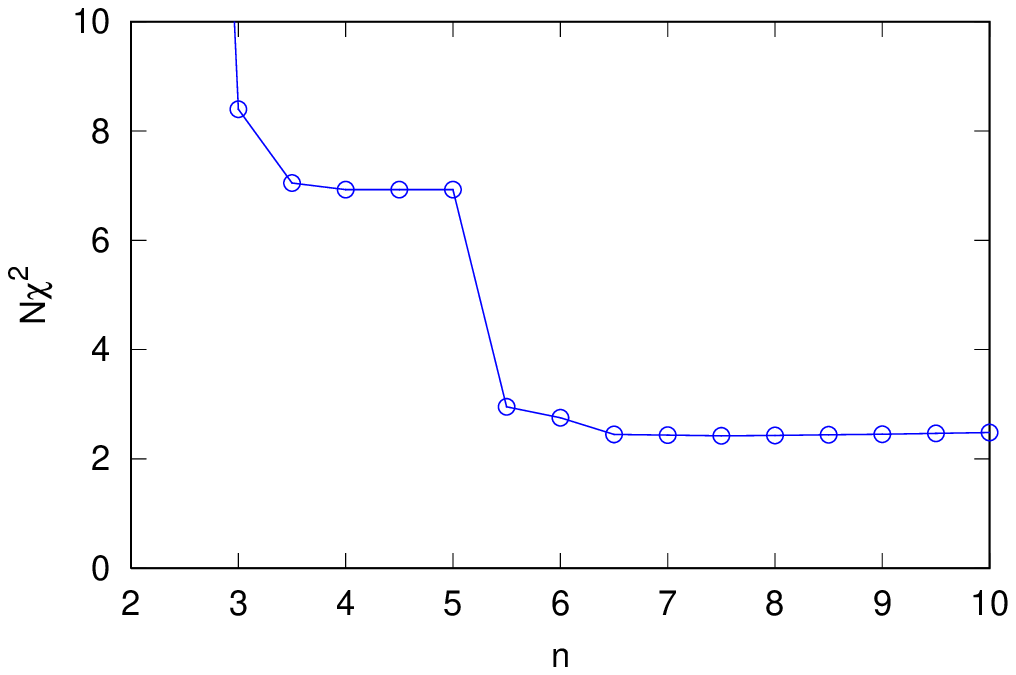}
\includegraphics[width=0.32\textwidth]{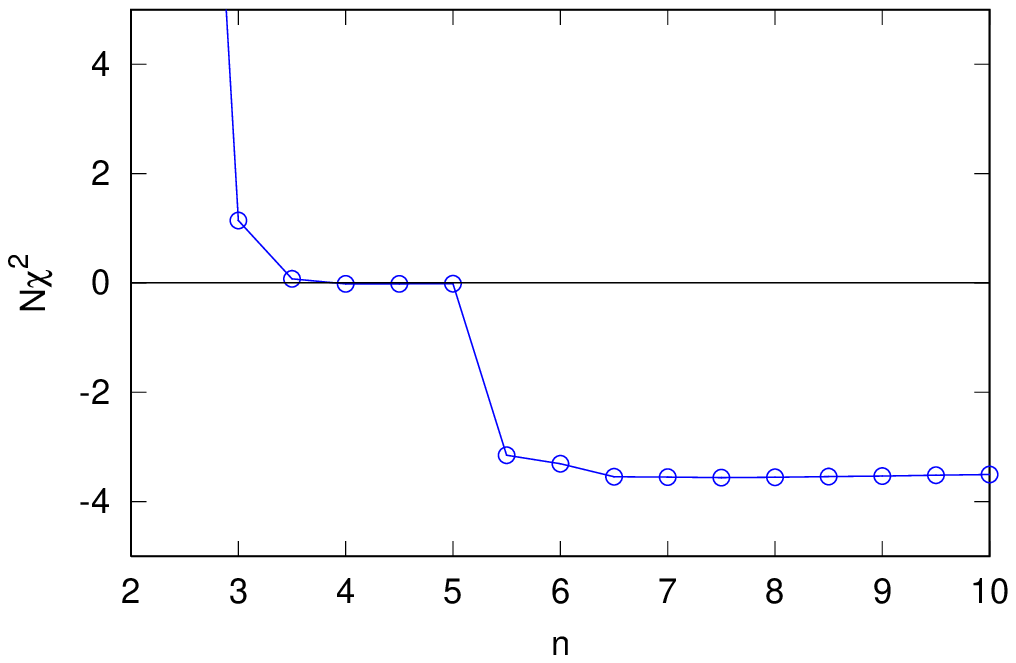}
\includegraphics[width=0.32\textwidth]{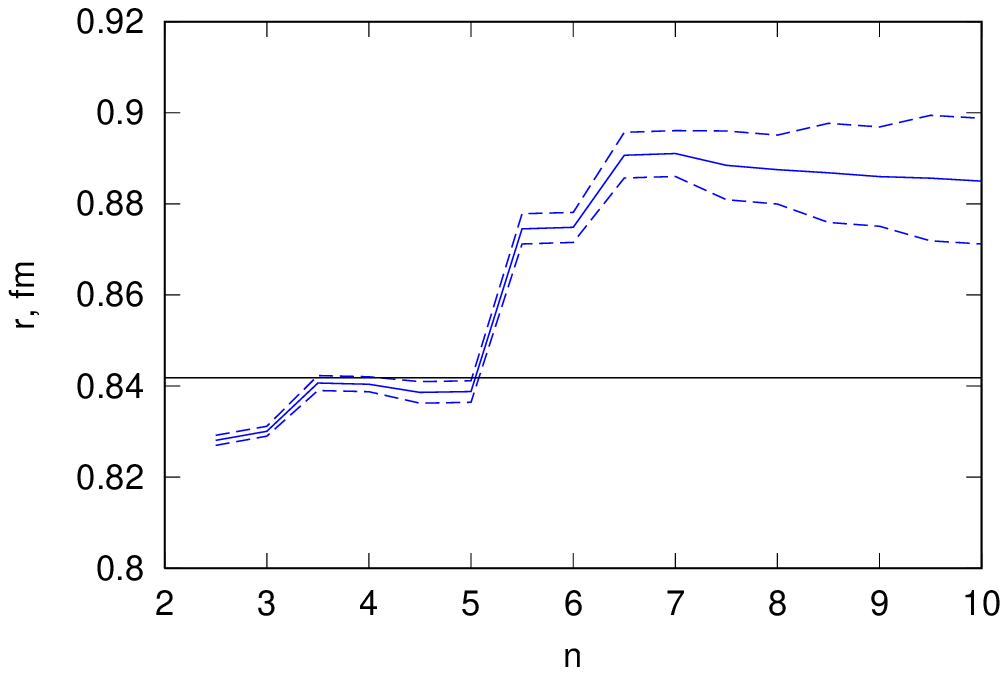}
\caption{Bernauer data fitting results at $Q_{max}^2 = 1 \GeV^2$ and different $n$. Left: $N\chi^2$. Center: same, but after increasing point-to-point errors by 12.5\%. Right: extracted radius (solid), with $\pm 1\sigma$ errors (dashed); horizontal line is muonic result.}
\label{fig:real}
\end{figure}

The curves, similar to Fig.~\ref{fig:pseudo}, but based on real experimental data from Ref.~\cite{Bernauer}, are shown in Fig.~\ref{fig:real}.
We see that the behaviour has a difference from pseudo-data results.
First, while it seems like $N\chi^2$ begin to stabilize at $n = 4..5$,
it abruptly drops at $n = 5.5$, and then continues at approximately constant level up to the end.
Second, its values are unusually large: first stabilization occurs at $N\chi^2 \approx 6.9$,
and the second stable value is about 2.4, not close to zero as expected from pseudo-data simulations.
Note that just at the point where $N\chi^2$ unusually drops, the extracted $r_E$ raises from $\sim 0.84$~fm to $\sim 0.88$~fm.

Large values of $N\chi^2$ may suggest that the experimental errors are a bit underestimated. Note that the errors estimation procedure of Ref.~\cite{Bernauer} (see also \cite{BernauerThesis}) is non-standard, rather complicated, and was furthermore criticized e.g. in Ref.~\cite{ArringtonCritics}.
Trying to renormalize experimental errors, we have two alternatives: to assume that real stabilization occurs either at $n>6$ or at $n = 4..5$.
In the first case, we need to enlarge errors only by 4.5\%, and the obtained $r_E$ will be about 0.88~fm, but the behavior of $N\chi^2$ at lower $n$ would be difficult to explain.

On the other hand, enlarging errors by 12.5\%, we obtain the picture in the center panel of Fig.~\ref{fig:real}.
The $N\chi^2$ begin to stabilize around zero level at $n = 4..5$. Then, according to pseudo-data results,
the radius obtained there should be close to true one. And actually, this value is very close to the muonic result --- see Fig.~\ref{fig:real}, right panel.
The abrupt drop of $N\chi^2$ at $n = 5.5$ may then indicate that there is some highly correlated component in the point-to-point errors (remember that in our analysis we silently assume that all errors except normalization are uncorrelated).
It is clear that in presence of such a correlated error the $\chi^2$ may drop, and the radius will deviate from the true one, since the fit will partly follow the error profile instead of true FF. So, this option may reconcile Mainz data and muonic hydrogen experiments.

\section{Conclusions}
We reanalyzed electron-proton scattering data in the momentum transfer region $Q^2 < 1\GeV^2$, which were used to determine proton electric radius $r_E$,
studying dependence of $\chi^2$ and resulting $r_E$ on the cut-off in momentum transfer $Q_{max}^2$ and fitting polynomial degree $n$, using both pseudo-data and real experimental data of Bernauer et al. \cite{Bernauer, BernauerThesis}.

The normalization coefficients are better determined at higher $Q_{max}^2$, and after they are fixed this way, the obtained $r_E$ does not depend on $Q_{max}^2$.

With pseudo-data, $\chi^2$ smoothly decreases with $n$, and when it becomes approximately constant, $r_E$ appears close to the true value.
When fitting Bernauer data, $\chi^2$ shows unusual dependence on $n$, and possible explanation is that at $n \ge 6$ some correlated point-to-point error comes into play, distorting obtained $r_E$ value. If this is the case, better approximation of true radius is obtained at smaller $n = 4..5$, and that value is consistent with the muonic hydrogen results.

\end{document}